\documentclass[final]{elsart3p}

\usepackage{graphicx}

 \journal{Comp. Mat. Sci.}
\usepackage{dcolumn}
\usepackage{bm}

\begin{document}

\begin{frontmatter}

\title{Electronic and gap properties of lead-free perfect and mixed
hybrid halide  perovskites: An \textit{ab-initio} study}

\author{Athanasios Koliogiorgos, Sotirios Baskoutas and Iosif Galanakis}
\address{Department of Materials Science, School of Natural
Sciences, University of Patras,  GR-26504 Patra, Greece}

\begin{abstract}
Hybrid halide perovskites are currently under intense
investigation due to their potential applications in
optoelectronics and solar cells. Among them, MAPbI$_3$ where MA
stands for the methylammonium cation, exhibits ideal properties
for solar cells. In attempt to identify new lead-free halide
perovskites we have studied using \textit{ab-initio} electronic
structure calculations in conjunction with hybrid functionals a
series of MABX$_3$ compounds where B is a divalent cation and X is
a halogen atom.  Our results suggest that the compounds under
study exhibit a variety of lattice constants and energy band gaps.
Especially, MAGeCl$_3$ and MAGeBr$_3$ are susceptible to replace
MAPbI$_3$ in devices since they show comparable energy gaps.
Further calculations on the mixed hybrid halide perovskites show
that we can tune the values of the energy gap although no
simplified pattern exists. Our results pave the way for further
investigation on the use of these materials in technology relevant
applications.
\end{abstract}

\begin{keyword}
Halide perovskites \sep Density-functional theory \sep Electronic
Band structure \PACS 71.20.-b \sep 71.20.Nr \sep 71.15.Mb
\end{keyword}

\end{frontmatter}

\maketitle

\section{Introduction}

Renewable energy sources are essentially a one-way road for energy
consumption in the 21st century, since the finite and limited
supply of fossil fuels is bound to come to an end. The sun is a
practically infinite energy source, and photovoltaic(PV)  devices
that turn the energy of sunlight into electricity are expected to
become more prominent as a main renewable energy source
\cite{Yang2016}. Silicon is the most widely used material in solar
cell technology, with a history of over 60 years \cite{Green1993}.
Other types of solar cells used in PV technology include type
III-V semiconductors, quantum dots, dye-sensitized solar cells,
organic solar cells and perovskites.

Designated as perovskite is a material that has the general
structure of ABX$_3$. The most known perovskites are the ones
where X is an oxygen. To achieve charge neutrality in that case, A
has to be a cation of +2 valence and B a cation of +4 valence of
dissimilar size like in CaTiO$_3$ \cite{Pena2001}. The versatility
of perovskites makes them highly attractive as they can form
multidimensional structures pertaining to the same chemical
formula through use of different combinations of various
components \cite{Chen2015,Gratzel2014}. A wide variety of elements
may be incorporated in the ABX$_3$ structure, as long as the
requirement for charge neutrality is satisfied. Although oxide
perovskites are the most well-studied and widely used in
applications due to their multifunctional nature, their wide band
gaps limit  their use in solar cell technology, as they utilize a
mere 8-20\%   of the solar spectrum \cite{Pena2001}.

To overcome the poor absorption of the oxygen perovskites, other
types have been proposed like the halide perovskites where the X
anion is a halogen instead of oxygen and mainly the so-called
hybrid or organometallic halide perovskites, where the A cation is
an organic molecule
\cite{Yang2016,Gratzel2014,Hoefler2017,Papavassiliou2012},
offering the ability to tune the photoconductive properties
through varying halide components \cite{Weber1978}. Their success
is based on  a highly favorable charge-carrier mobility so that
both light absorption and charge conduction are possible
\cite{Hohnston2015}. One of the most widely used organic cations
is methylammonium (MA) which has the chemical formula
CH$_3$NH$_3$$^+$ \cite{Yang2016}. Especially the one containing
lead as the divalent cation and iodine as the halogen, MAPbI$_3$,
has attracted most of the attention
\cite{Brittman2015,Frost2014,Filippetti2014}, and  its growth
conditions have been widely studied \cite{Albero2016,Zhou2016}.

MAPbI$_3$ has an experimental gap of 1.5 eV, which makes it
suitable for absorption in the optical regime \cite{Zhao2015}.
More recent experimental results on cubic crystals of MAPbI$_3$
gave a value of 1.69 eV at 330 K \cite{Quarti2016}. Although
several of its properties and their effect on the band gap have
been widely studied, such as the influence of the orientation of
the MA atoms \cite{Brivio2013,Walsh2015,Leguy2015}, the toxicity
of the lead atoms led to the search for alternative hybrid halide
perovskites \cite{Stoumpos2014}. \textit{Ab-initio} electronic
structure calculations have been also employed in this search. Sn,
which is isovalent to Pb, has been proposed as a replacement but
it oxidizes very easily and the perovskite structure is destroyed
\cite{Bernal2014,Borriello2003}. Sr replacement for Pb in
MAPbI$_3$ leads to an energy gap double the initial one and thus
cannot be applied in PV applications \cite{Jacobsson2015}. The
partial substitution of I atoms with other halogen influences the
energy width of the band gap but does not solve the toxicity
problem \cite{Mosconi2013,Motta2015}.

Motivated by the search for lead-free hybrid halide perovskites,
we carry out an extended \textit{ab-initio} study of the MABX$_3$
compounds. As X we have considered all possible halogen atoms,
namely F, Cl, Br and I. Divalent B cations can be either the
alkali earth elements (Ca, Sr, Ba) since they have two valence
\textit{p} electrons, the late transition metal atoms (Zn, Cd, Hg)
which have two valence \textit{s} electrons and the metalloids
(Ge, Sn, Pb) which have also two valence \textit{p} electrons but
contrary to the alkali earth elements the valence \textit{d}
states are completely occupied. For all 36 resulting compounds we
have relaxed the position of the atoms and the C-N bond length of
the MA cation and determined the equilibrium lattice constants.
For the equilibrium lattice constants, we have employed advanced
functionals of the exchange and correlation energies and have
determined the width of the energy band gaps. Finally, we have
also studied the possibility to tune the energy band gaps in mixed
hybrid halide perovskites by mixing two kinds of halogen atoms in
the unit cell. In section 2 we present the details of our
calculations. Section 3 is devoted to the structural properties of
the compounds under study and section 4 to their electronic and
gap properties. In section 5 we present our results on the mixed
compounds and finally in section 6 we summarize and present our
conclusions.

\begin{figure}
\includegraphics[width=\columnwidth]{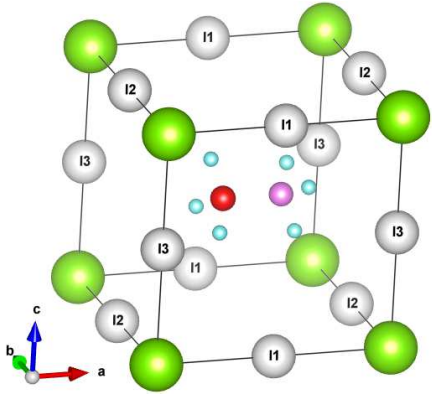}
\caption{Cubic unit cell of the MABI$_3$ compounds. The divalent
cations B (green spheres) are at the corners of the cube
surrounded by six halogen atoms (grey atoms). There are three
inequivalent halogen atoms in the unit cell at the midpoints of
the edges. The C-N bond of the MA cation (CH$_3$NH$_3$) is along
the x-axis, and each one of the C (red sphere) and N (purple
sphere) atoms is surrounded by three H atoms (light blue spheres).
We also denote the Cartesian axis $x$, $y$ and $z$ using the unit
vectors $\vec{a}$, $\vec{b}$ and $\vec{c}$, respectively. }
\label{fig1}
\end{figure}

\section{Computational method}

We have calculated the structural and electronic properties of the
hybrid halide perovskites under study  employing the
\textit{ab-initio} total energy and molecular-dynamics program
VASP (Vienna \textit{Ab-initio} Simulation Package) developed at
the Institut f\"ur Metaliphysik of the Universit\"at Wien
\cite{VASP}. We made use of the projector augmented waves (PAW),
which are a generalization of the ultrasoft pseudopotentials used
in conventional pseudopotential electronic structure methods
\cite{PAW}. To account for the exchange-correlation potential we
have used the generalized gradient approximation (GGA) as
formulated by Perdew, Burke and Ernzerhof (PBE) \cite{PBE}. More
precisely, we have  employed the parametrization of PBE developed
by Perdew et al. in reference \cite{PBEsol} which restores the
correct density-gradient expansion for the exchange energy in
solids with respect to the initial PBE formalism (known in
literature as PBEsol).

Although GGA in general is known to reproduce accurately the
structural properties, it presents a serious drawback regarding
semiconductor materials: while the calculated band structure is
qualitatively correct, the energy gap is largely underestimated.
But the exact value of the energy gap is crucial for applications
since it determines the wavelength of the absorbed light. To
restore the correct value of the energy gap several methods have
been proposed. One of the most rigorous is the use of more complex
functionals where the exchange energy is a mixture of the GGA and
Hartree-Fock functionals. These functionals are known as ``hybrid
functionals" and the most well-known in materials science is the
Heyd-Scuseria-Ernzerhof (HSE06) functional \cite{HSE06}, which has
been successfully implemented in VASP \cite{VASP-HSE06}. Except
HSE06, we have also performed calculations using the modified
Becke-Johnson functional in conjunction with the PBEsol one (known
as mBJ+PBEsol). The mBJ has been developed in 2006 by Becke and
Johnson in an attempt to provide an efficient exchange functional
which would reach the accuracy of the hybrid functionals like
HSE06 but which would need similar CPU resources like GGA
\cite{Becke} contrary to HSE06 based calculations which are very
demanding. It became popular in 2009 when Tran and Blaha
introduced it in the full-potential (linearized) augmented
planewave and local orbitals [FP-(L)APW+lo] method and found that
it gives band gaps for a series of insulators and semiconductors
close to the HSE functionals \cite{Tran2009}. We should also note
that the mBJ functional, actually, is a potential-only functional
being a local approximation to an atomic exact-exchange potential
plus a screening term which is used in conjunction to one of the
exchange-correlation schemes (PBEsol in our case). Thus the
mBJ-based calculations within VASP are not self-consistent with
respect to the total energy contrary to the HSE06 or PBEsol
functionals (details are given in the online manual of VASP
\cite{manual-vasp}).

We should also shortly discuss the accuracy of the various
functionals to study the compounds of interest. Although
calculations using HSE06 are expected to give more accurate
results than the usual GGA functionals and mBJ calculations are
expected to reach the accuracy of HSE06, in reality the success of
the HSE06 and mBJ functionals is materials specific and it is
established only when the the computed values are compared to the
experimental ones. HSE06 has been applied with success to the
study of the electronic properties of the oxygen perovskites
\cite{Franchini2014}. Moreover, recent experiments on cubic
crystals of MAPbI$_3$ by Quarti and collaborators produced a value
of 1.69 eV at 330 K \cite{Quarti2016}. As will be discussed in
section IV, PBEsol produces a value of 1.49 eV, while HSE06 and
mBJ+PBEsol produce values of 1.82 eV and 1.85 eV, respectively.
Since ab-initio calculations are carried out at 0 K and the width
of the band gap drops with the temperature, we expect that our
calculated HSE06 and mBJ+PBEsol band gap values are close to the
experimental one. All the above provide strong evidence that HSE06
and mBJ+PBEsol are suitable to study the halide perovskites,
although more experimental results on crystals are needed for
definitive conclusions.

Concerning the details of the calculations, we have used for all
of them a cutoff for the kinetic energy of the plane waves of 500
eV and for the Ge, Sn and Pb atoms we have included in the PAW
basis the 3\textit{d}, 4\textit{d} and 5\textit{d} orbitals,
respectively, as valence states. For the case of the mBJ+PBEsol
calculations we have used a more advanced basis set including also
the kinetic energy density of the core electrons. In order to
establish the accuracy of our calculations we have calculated
MAPbI$_3$ as a test case, since it is the most studied compound in
literature. The obtained results were in perfect agreement with
the existing ones in literature \cite{Brivio2013}, as will be
discussed later.

\begin{figure}
\includegraphics[width=\columnwidth]{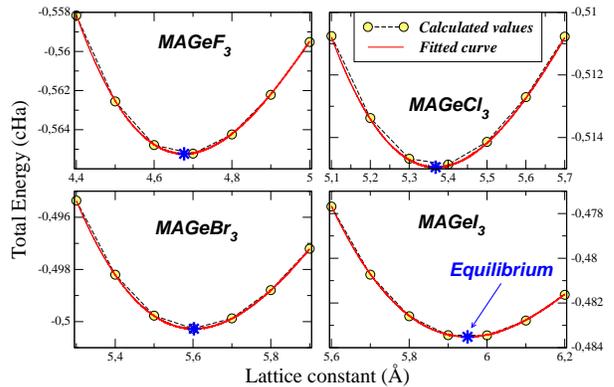}
\caption{Calculated total energy versus the lattice constant in
MAGeX$_3$ compounds. The calculated values have been fitted with a
third order polynomial in order to determine the equilibrium
lattice constant. Note that the use of the Murnaghan equation of
state gives identical results in the case of the compounds under
study. } \label{fig2}
\end{figure}

\section{Structural properties}

The hybrid halide perovskites under study crystallize in a cubic
structure shown in figure \ref{fig1} \cite{Leguy2015}. The
divalent cations sit at the corners of the cube surrounded by
halogen atoms in an octahedral environment. The MA cation sits at
the center of the cube. The first step in our study is the
determination of the equilibrium lattice constant. To determine it
for all the compounds we have performed total energy calculations
for several lattice constants using a 6$\times$6$\times$6
Monkhorst-Pack grid in the 1st Brillouin zone \cite{Monkhorst} in
conjunction with the PBEsol functional. For each lattice constant
we fixed the position of the divalent cation and the halogen
atoms. We allowed the MA atom to fully relax within the cell (the
positions of all C, N and H atoms have been allowed to change).
The only constraint, which was imposed, was that the C-N bond
stays parallel to the [100] direction passing from the center of
the cube. In reality as shown in reference \cite{Leguy2015} the MA
cations can rotate in very short times but such a picture cannot
be captured by conventional electronic structure calculations.
Moreover, test calculations for MAPbI$_3$ have shown that the
orientation of the MA cation is not decisive for the obtained
values of the band gap. We have chosen 7 values around the
equilibrium and fitted a third order polynomial curve to determine
the equilibrium lattice constant corresponding to the minimum of
the energy. In figure \ref{fig2} we show the calculated total
energy values versus the lattice constants as well as the fitted
curve for the four MAGeX$_3$ compounds. Although the curve has
different curvature for each of these compounds, the fitting is
excellent and the fitted curve passes in all cases through all
seven calculated points.

\begin{table}
  \caption{Calculated equilibrium lattice constants, $a_{eq}$, in \AA\ for the hybrid halide
  perovskites  using the PBEsol approximation. These perovskites have the chemical formula
  MABX$_3$ where MA is the methylammonium cation (CH$_3$NH$_3$), B is a divalent
  cation and X is a halogen atom.}
  \label{table1}
  \begin{tabular}{lcccc}
    \hline

$a_{eq}^\mathrm{MABX_3}$(\AA )&  X=F &  X=Cl&  X=Br & X=I \\
\hline

B=Ca & 4.7059 & 5.4883 & 5.7590 & 6.1709\\

B=Sr & 4.8907 & 5.7103 & 5.9886 & 6.4127\\

B=Ba & 5.1694 & 6.0142 &6.3127 &6.7362\\

B=Zn & 4.4958 & 5.1723 & 5.4276& 5.8133\\

B=Cd & 4.6656 & 5.3548 &5.6088 &5.9923\\

B=Hg & 4.7385 & 5.4047 &5.6563  &6.0203\\

B=Ge & 4.6771 &5.3679  &5.6023 &5.9499\\

B=Sn & 4.8581 &5.6047  &5.8431  &6.2021\\

B=Pb &4.9354 & 5.7006  &5.9454 &6.3134\\ \hline
  \end{tabular}
\end{table}

We have gathered the calculated equilibrium lattice constants in
\AA\ for all 36 calculated compounds and present them in table
\ref{table1}. The obtained equilibrium lattice constants scan a
wide range of values starting from 4.4958 \AA\ in the case of
MAZnF$_3$ up to 6.7362 \AA\ for MABaI$_3$. As expected, the trends
follow the size of the atomic radius of the constituent atoms. As
we move from the light F halogen to the heavy I atom, keeping the
divalent cation B constant, the equilibrium lattice constant
increases in all cases by about 30\%. Differences are much smaller
if we fix the halogen atom and  change the B cation within the
same row. We can compare our calculated values with previous
calculations and experiments in the case of MAPbI$_3$ for which
data exist in literature. Powder diffraction experiments gave a
value of 6.26 \AA\ \cite{Baikie2013}, while calculations give
values between 6.26 \AA\ and 6.33 \AA\
\cite{Brivio2013,Mosconi2013}, which are very close to our
calculated value of about 6.31 \AA .

\begin{figure}
\includegraphics[width=\columnwidth]{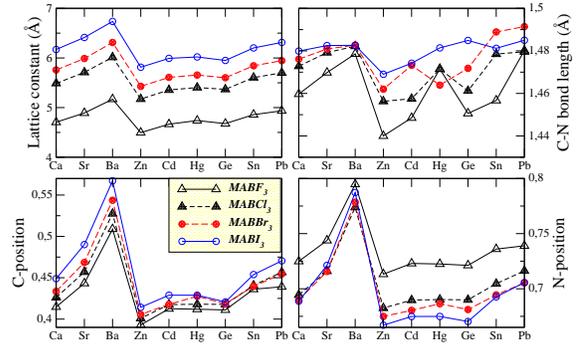}
\caption{Behavior of the equilibrium lattice constant (upper left
panel), C-N bond length (upper right panel) in \AA , and the
position of the C and N atoms of the MA cation (lower panels) in
units of the lattice constant as a function of the divalent
cation. } \label{fig3}
\end{figure}

It is interesting also to study the behavior of the MA cation in
the studied compounds. In figure \ref{fig3} we have gathered all
the related results as a function of the divalent B cation. The
four curves in each panel correspond to the four halogen atoms
under study. In the upper left panel, we first present the
behavior of the equilibrium lattice constants. All four curves are
parallel and thus for each divalent cation B, the halogen atoms
affect the lattice constant in the same manner. In the upper right
panel we present the behavior of the C-N bond length. The latter
in general follows the behavior of the equilibrium lattice
constants but there are exceptions, \textit{e.g.} the length is
larger for MAPbBr$_3$ than MAPbI$_3$. The bond length does not
show large variations being the smallest for MAZnF$_3$ (about 1.44
\AA ) and the largest for MAPbB$_3$ (about 1.49 \AA ), a variation
of about 3.5\%  only. In the lower panels in figure \ref{fig3} we
present the position of the C and N atoms as a fraction of the
lattice constant. First, we have to note that in all cases the MA
cation is not centered around the middle of the cube resulting in
the appearance of a dipole where the N atom is more distant from
the cube's center. The Ba-based compounds are an exceptional case
since the MA  cation is completely off-center and both C and N
atoms are on the right side of the cube's center. Thus, it is
possible to find ferroelectric behavior in these compounds.
Finally, we should note that, although all curves for the
different structural properties in figure \ref{fig3} follow the
shape of the lattice constant curves in the upper left panel, we
were not able to deduce some strict rules as is the case in other
phenomena \cite{Pei2015}.

\section{Electronic and gap properties}

Since we have determined the equilibrium lattice constants, we
proceeded with the calculation of the energy gaps which is also
the main finding of the present study. We have used a much denser
10$\times$10$\times$10 Monkhorst-Pack grid to carry out the
self-consistent electronic band structure calculation at the
equilibrium lattice constants. First, we employed the PBEsol
functional. As mentioned above, the former is not accurate enough
in most cases to compute the energy gaps, and thus we used the
PBEsol calculated electronic charge and wavefunctions as the input
to perform electronic band structure calculations with the most
accurate and much more demanding in computer resources hybrid
HSE06 functional as well as the more efficient mBJ+PBEsol using
the same grid in the reciprocal space as for the PBEsol
calculations. In table \ref{table2} we have gathered the HSE06
calculated energy gaps for all 36 six compounds under study as
well as the mBJ+PBEsol and PBEsol results in braces and
parentheses, respectively. In all cases under study, the use of
HSE06 and mBJ+PBEsol leads to larger values for the band gaps with
respect to the PBEsol functional. All three functionals give the
same trends within the table as we change either the divalent B
cations or the X halogen atom, but the effect of HSE06 with
respect to PBEsol is not uniform in all cases. \textit{E.g.} for
MAZnF$_3$ the HSE06 gap is almost two times the PBEsol gap while
in MACaF$_3$ the increase is only 25\%. Also the behavior of
mBJ+PBEsol is not uniform in all cases. In almost all cases
mBJ+PBEsol gives band gaps smaller than the HSE ones, but there
are compounds like MACaCl$_3$ where the mBJ+PBEsol is closer to
the PBEsol than the HSE06 value, and there are compounds like
MAGeI$_3$ and MAPbI$_3$ where both HSE06 and mBJ+PBEsol produce
identical values. Interestingly, in the case of the
gapless-semiconducting compounds like MAZnI$_3$ all three
functional give identical results. Thus, the effect of the HSE06
and mBJ+PBEsol functionals on the calculated energy gap values is
materials specific.

\begin{table*}
\caption{Calculated energy gap in eV using  the HSE06, the
mBJ+PBEsol [in braces], and PBEsol functionals (the latter in
parentheses) for the MABX$_3$ hybrid halide perovskites. The zero
values correspond to a gapless (zero-gap) semiconducting
behavior.}
  \label{table2}
  \begin{tabular}{lc|c|c|c}
    \hline

& \multicolumn{4}{c}{Band gap (eV)} \\
& \multicolumn{4}{c}{HSE06 [mBJ+PBEsol] (PBEsol)} \\
MABX$_3$  &  X=F &  X=Cl&  X=Br & X=I \\
\hline

B=Ca &7.71[7.30](6.20)  & 6.30[5.20](4.80) & 5.36[4.63](4.30) & 4.00[3.57](3.40) \\
B=Sr &7.57[7.28](5.92)  & 6.30[5.70](4.92) & 5.26[5.08](4.15) & 4.23[4.00](3.52) \\
B=Ba &6.91[6.53](5.33)  & 5.84[5.30](4.89) & 5.02[4.64](4.26) & 3.86[3.84](3.27) \\
B=Zn &3.78[3.35](1.96)  & 1.94[1.80](0.81) & 0.71[0.84](0.12) & 0.00[0.00](0.00) \\
B=Cd &4.47[4.05](2.90)  & 2.42[2.11](1.52) & 1.30[1.21](0.57) & 0.22[0.23](0.00) \\
B=Hg &1.84[1.62](0.70)  & 0.23[0.00](0.00) & 0.00[0.00](0.00) & 0.00[0.00](0.00) \\
B=Ge &4.12[3.33](2.66)  & 1.96[1.64](1.33) & 1.51[1.37](1.07) & 1.21[1.22](0.93) \\
B=Sn &3.61[3.07](2.24)  & 1.70[1.55](1.07) & 1.19[1.10](0.84) & 0.94[0.86](0.72) \\
B=Pb &4.78[4.16](3.28)  & 2.91[2.52](2.24) & 2.44[2.28](1.73) & 1.82[1.85](1.49) \\
\hline
  \end{tabular}
\end{table*}

\begin{figure}
\includegraphics[scale=0.4]{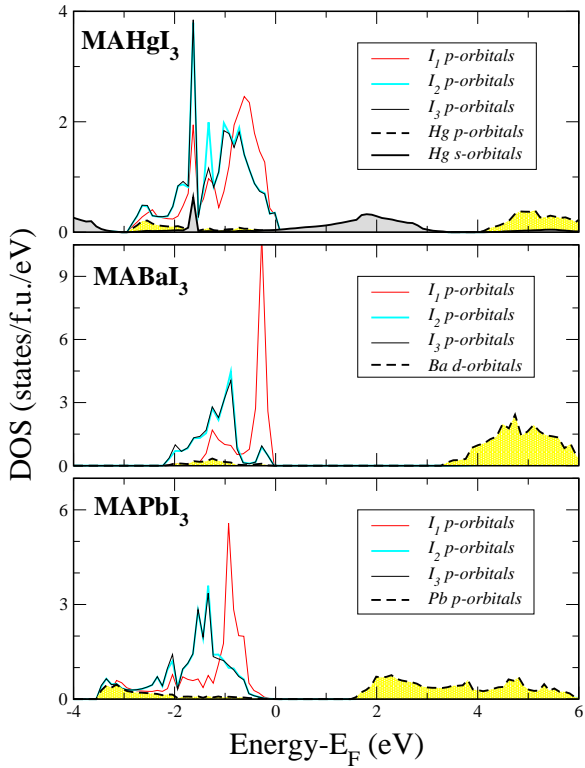}
\caption{Orbital and atom-resolved density of states (DOS) as a
function of the energy for MA(Hg,Ba,Pb)I$_3$ compounds using the
PBEsol functional. The zero energy corresponds to the Fermi level.
The valence bands consist of the I \emph{p}-orbitals. The
conduction bands are made up from the Ba, Hg or Pb  states. The
kind of hybridization determines the width of the energy gap. Note
that the use of HSE06 or mBJ+PBEsol functional opens the gap but
does not influences the character of the bands.} \label{fig4}
\end{figure}

For all cases of the divalent B cations, the energy band gap
becomes considerably smaller as we change the halogen atom going
from the light F to the heavy I one, with the largest variation
being observed in the cases of the Zn, Cd and Hg cations. This is
expected since the lighter the halogen atom is, the deeper are its
valence \textit{p} states and the larger is the energy gap,
\textit{i.e.}, in the case of F the valence states are the
2\textit{p} orbitals while in the case of I the valence states are
the 5\textit{p} orbitals which are much higher in energy (see
figure \ref{fig4}). Our calculated values are in agreement with
previous calculations where available. More precisely the energy
gap for MASrI$_3$ using PBEsol was estimated to be 3.6 eV
\cite{Jacobsson2015},  very close to our value of 3.52 eV. For
MAPbI$_3$, values of 1.5 eV using PBEsol and about 2.0 eV using
the HSE06 functional have been obtained \cite{Brivio2013}. Our
PBEsol value, as shown in table \ref{table2}, is 1.49 eV,
identical to the one in literature, while our HSE06 value is 1.82
eV -slightly smaller than the 2 eV in literature.

We can categorize the materials presented in table \ref{table2} in
three categories depending on the B cation. When B is an alkali
earth atom (Ca, Sr or Ba) the calculated energy gaps are very
large exceeding the 4 eV, with the exception of MABaI$_3$ where it
is about 3.9 eV using both HSE06 and mBJ+PBEsol. For the case of
X=F they are even close to 7 eV. These values are very large and
by far exceed the values of 1.5-1.8 eV required in order to absorb
in the optical regime. As a consequence, these materials are
transparent to the visible light. When the cation is Zn, Cd or Hg
energy gaps are considerably smaller. There is the case of
MAZnCl$_3$ and MACdBr$_3$ where the gap is 1.94/1.80 eV  and
1.30/1.21 eV, respectively, using the HSE06/mBJ+PBEsol
functionals, and thus could be useful for applications, although
Cd is a toxic material like Pb. But it is also widely used in
compounds like, \textit{e.g.}, the CdSe quantum dots for LED
applications. Some very interesting cases are all three compounds
with I, as well as MAHgBr$_3$ where the energy gap is very small
or zero. Such nearly-gapless or gapless semiconductors are
interesting for spintronic applications due to the high mobility
of carriers and the very small amount of energy needed to excite
the electrons. Finally, we have the case where the cation is one
of the metalloids Ge, Sn or Pb. Sn halide perovskites have been
suggested in literature as a possible substitute for the Pb ones,
but they are unstable and the Sn atoms are oxidized very easily
going from a +2 to a +4 state, destroying the perovskite structure
\cite{Borriello2003}. Thus the case of the Ge perovskites becomes
much more interesting. The Ge based halide perovskites show band
gaps slightly smaller than the Pb-based ones.  Both MAGeCl$_3$ and
MAGeBr$_3$ with band gaps of 1.96 eV and 1.51 eV within HSE06
(mBJ+PBEsol gives slightly smaller values), respectively, are
susceptible of absorbing light in the visible region and thus
could be employed in solar cell applications instead of MAPbI$_3$
used in current solar cells.

To elucidate the trends observed in the calculated bang gap
energies,  we have plotted in figure \ref{fig4} the atom- and
orbital resolved density of states (DOS) for three compounds
MA(Hg, Ba or Pb)I$_3$ using the PBEsol functional; HSE06 and
mBJ+PBEsol functionals affect the width of the band gap but not
the character of the bands. In all three cases the valence bands
are made up from the \textit{p} states of the iodine atoms. There
is also a very small \textit{s} admixture which is not shown here.
Note that there are three inequivalent I atoms denoted as I$_1$,
I$_2$ and I$_3$ (for the definition please refer to figure
\ref{fig1}). I$_2$ and I$_3$ are almost equivalent with respect to
the position of the MA cation and thus have very similar DOS which
cannot be distinguished. On the contrary, I$_1$ is located in the
middle of the [100] edge and has a distinctively different DOS
with its \textit{p}-states occupying the higher part of the
valence bands. In the case of MAPbI$_3$ the conduction band is
made up from the Pb empty 6\textit{p} states and thus the gap is
due to the \textit{p}-\textit{p} hybridization, in agreement with
previous \textit{ab-initio} results \cite{Filippetti2014}. Similar
is the situation when instead of Pb we have the isovalent Sn or Ge
cations. In the case of alkali earth cations the conduction band
is now made up from the unoccupied \textit{d} orbitals (the
unoccupied  \textit{p} states are higher in energy) and more
precisely the triple degenerate $t_{2g}$ orbitals which transform
in the same way as the \textit{p} orbitals in the case of
tetrahedral and octahedral symmetries, \textit{i.e.} in the case
of MABaI$_3$ shown in figure \ref{fig4} the conduction band
consists of the Ba 5\textit{d} orbtials and the 6\textit{p}
orbitals are higher in energy. Therefore the main interaction
responsible for opening the gap is the \textit{p}-\textit{d}
hybridization which is not as strong as the \textit{p}-\textit{p}
one and the gap is larger. Finally, when the cation is Zn, Cd or
Hg, a conduction \textit{s}-state appears clearly between the
occupied \textit{p} states of the halogen atoms and the unoccupied
\textit{p} states of the Zn, Cd or Hg atoms. Thus although the
\textit{p}-\textit{p} hybridization opens a sizeable gap, the
location of the Zn, Cd, or Hg \textit{s}-states within the gap
leads to much smaller energy gap values and even to the observed
gapless  behavior as in the case of the MAHgI$_3$ compounds shown
in the upper panel of figure \ref{fig4}.

\begin{table*}
  \caption{Calculated energy gaps in eV using the PBEsol functional
  for the mixed hybrid halide perovskites.
  We have assumed linear variation of the lattice constants between the extreme perfect compounds.
  Note that the terms ``I, II and III" correspond to the position of the single halogen atom in the unit
  cell (see text for definitions). We have also calculated the energy gaps for the perfect compounds
  (last two lines in each case) at the
  lattice constants of the mixed hybrid halide perovskites.}
  \label{table3}
  \begin{tabular}{lcc|lcc}
    \hline

  &  MAGe(Cl$_2$I)   & MAGe(I$_2$Cl) &  &  MACa(Cl$_2$I)    &
  MACa(I$_2$Cl)\\ \hline

I-case   & 0.93 & 0.70 &  I-case   & 4.01 & 3.48 \\

II-case  & 0.71 & 0.82  & II-case  & 4.17 & 3.51 \\

III-case & 0.83 & 0.80  & III-case & 4.01 & 3.48 \\

MAGeCl$_3$ &  1.58 &   1.82  &  MACaCl$_3$ &4.75& 4.25 \\

MAGeI$_3$ & 0.69 & 0.81 &   MACaI$_3$ & 2.98 & 3.16 \\  \hline

&&&&&\\

  &  MAPb(Cl$_2$I) &     MAPb(I$_2$Cl)    & &   MABa(Cl$_2$I) &
  MABa(I$_2$Cl)\\ \hline

I-case &1.41   &  1.51  &  I-case &4.04 &3.66\\

II-case& 1.52  &  1.40  &  II-case &4.00 &3.65\\

III-case& 1.61 &  1.30  &  III-case &4.04 &3.62\\

MAPbCl$_3$ &2.44  &  2.52  &  MABaCl$_3$ &4.58 &4.18\\

MAPbI$_3$  &1.01  &  1.30  &  MABaI$_3$  &3.34& 3.26\\ \hline

&&&&&\\

&    MAPb(Cl$_2$Br) &  MAPb(Br$_2$Cl) & &  MABa(Cl$_2$Br)&    MABa(Br$_2$Cl)\\

I-case & 2.04    &    1.83  &  I-case & 4.64 &4.27\\

II-case& 1.84     &   1.93  &  II-case&  4.54 &4.30\\

III-case  &1.94      &  1.83 &   III-case  & 4.55 &4.30\\

MAPbCl$_3$ &2.25&  2.44   &    MABaCl$_3$ &4.69 &4.67\\

MAPbBr$_3$ &1.53&   1.73 &      MABaBr$_3$ &4.36& 4.25\\

 \hline
  \end{tabular}
\end{table*}

\section{Mixed halide perovskites}

In the last part of our study we have concentrated on the
so-called mixed hybrid halide perovskites \cite{Mosconi2013}. The
idea is to seek whether by mixing the halogen atoms, one could
result in tuning the energy gap. We have employed the PBEsol
formalism since, as discussed in the previous section, PBEsol
accurately reproduces the correct trend with respect to the other
two more elaborated functionals. We have assumed for all the
studied compounds that the lattice constant varies linearly
between the equilibrium lattice constants. Thus for the
MAB(X$_2$X') compound the lattice constant is
\begin{equation} \label{eq}
a_\mathrm{MAB(X_2X')}=\frac{2}{3}a_\mathrm{MABX_3}+\frac{1}{3}
a_\mathrm{MABX'_3}.
\end{equation}
To validate our assumption, we have also computed the equilibrium
lattice constant using total energy calculations as for the
perfect compounds in the case of the Pb-mixed halide perovskites;
the calculated values differ less than 0.02 \AA\ from the values
deduced using equation \ref{eq}. Thus the deviation is less than
0.3 \% , which has negligible effect on the calculated electronic
properties, and we can safely use equation \ref{eq}.

For all studied mixed compounds,  the MA cation was relaxed
similarly to the ordered compounds described above. In the case of
the MAB(X$_2$X') there are three different configurations
depending on the position of the X' atom. We denote them as cases
I, II and III depending on whether the X' atoms substitute the X1,
X2 or X3 atom, respectively (this notation follows the
nomenclature introduced in figure \ref{fig1} for the I atoms).
Also, for each lattice constant we have recalculated the energy
gap for both MABX$_3$ and MABX'$_3$ parent compounds since the
value of the energy gap is lattice constant dependent and differs
from the value at the equilibrium lattice constant calculated and
presented in the previous section.

We have gathered the calculated energy band gaps for the mixed
hybrid halide perovskites in table  \ref{table3}. First, we have
to remark that the energy gap for the same compound can differ
substantially with the lattice constant. \textit{E.g.} MAPbI$_3$
has a gap of 1.01 eV at the lattice constant of MAPb(Cl$_2$I),
1.30 eV at the lattice constant of MAPb(I$_2$Cl) while at its own
lattice constant it is 1.49 eV as shown in table \ref{table2}.
Moreover, the energy gap varies in most cases less than 0.1 eV
 within the three possible cases concerning the position
of the X' atom, with the maximum of 0.2 eV attended for the
MAPb(Cl$_2$I) and MAPb(I$_2$Cl) compounds. Thus the exact
arrangement of the X and X' halogen atoms is not crucial for the
energy gap.

If we concentrate on the MAGe(Pb)(Cl$_2$I) and MAGe(Pb)(I$_2$Cl)
compounds, the energy gaps of the mixed compounds are much closer
to the values for the perfect MAGe(Pb)I$_3$ compound rather than
those for the MAGe(Pb)Cl$_3$ compound, and the difference in the
gap size with the latter can be as large as 1 eV, even in the case
when we have two Cl and one I atoms.  Thus, the change of the
energy gap between the two parent compounds does not present a
linear behavior,  and the presence of iodine is predominant in
determining the energy gap. When instead of I there is Br as in
the MAPb(Cl$_2$Br) and MAPb(Br$_2$Cl) compounds, the behavior of
the MAPb(X$_2$X') energy gap changes and it is much closer to the
perfect compound, with the three X atoms approaching a linear
behavior. When instead of Ge(Pb) there is an alkali earth atom
like Ca or Ba, the energy gap shows an almost linear behavior with
the concentration of the halogen atoms between the two extreme
parent compounds, even in the case where iodine is part of the
halogen atoms contrary to the behavior of the MAGe(Pb)(Cl$_2$I)
and MAGe(Pb)(I$_2$Cl) compounds.

Consequently, taking into account the fact  that the energy gap of
the perfect compounds depends strongly on the lattice constant,
there is no simplified way to tune the properties of the mixed
hybrid halid perovskites, the property is materials specific and
should be determined in each case using electronic structure
calculations.

\section{Summary and conclusions}

Perovskites are currently under intense investigation due to their
potential applications in several devices ranging from solar cells
to LEDs. Hybrid halide perovskites are of special interest
combining the presence of an organic cation like methylammonium
(MA), having the chemical formula CH$_3$NH$_3$, with the presence
of halogen atoms. Especially MAPbI$_3$ has shown excellent
properties regarding its usage in solar cell applications.
However, the presence of lead prohibits it from a wide-spread use
in industrial applications.

In an attempt to identify new hybrid halide perovskites which
could substitute MAPbI$_3$, we have studied using
\textit{ab-initio} electronic structure calculations the
properties of MABX$_3$ compounds with the divalent cation B being
one of the Ca, Sr, Ba, Zn, Cd, Hg, Ge, Sn or Pb atoms and
considered all possible halogen atoms  X= F, Cl, Br and I.  The
first step in our study was to use the VASP \textit{ab-initio}
technique in conjunction with the so-called PBEsol functional for
the exchange-correlation to determine the equilibrium lattice
constants for all 36 compounds allowing the MA cation to relax
within the cubic unit cell. We then proceeded to employ the more
sophisticated hybrid HSE06 and  mBJ+PBEsol functionals to
calculate the electronic properties at the equilibrium and to
estimate the band gap. We have shown that both lattice constants
and energy gaps vary greatly depending on the choice for the
divalent cation and the halogen atom. Compounds based on alkali
earth atoms show very large band gaps, being transparent to light
in the optical regime, while the compounds based on Zn, Cd and Hg
show much smaller band gaps and some of them are even gapless
semiconductors. Among them, MAZnCl$_3$ and MACdBr$_3$ are
potential candidates to replace MAPbI$_3$. Due to deterioration of
the properties of the Sn-based compounds due to oxidation, the Ge
compounds seem to be very promising and, more especially, our
calculations suggest that MAGeCl$_3$ and MAGeBr$_3$ are suitable
to replace MAPbI$_3$ in devices. The width of the energy gaps
depends on the kind of hybridization forming the gap. We have also
performed calculations for the mixed hybrid halide perovskites
mixing two kind of halogen atoms. Although we could tune the
values for the energy band gaps, there is no simple pattern to
predict the exact value based on the properties of the parent
perfect compounds.

Consequently, we conclude that it is possible to find lead-free
perovskites that can be used in energy technology applications
like solar cells and optoelectronics, though the properties are
materials specific and extended state-of-the-art, heavy
\textit{ab-initio} calculations are a prerequisite for any
reliable result. We expect our results to intrigue further
experimental studies on these compounds since, even if some of the
calculated systems are ideal for a specific application, their
growth and incorporation in devices still requires advanced
synthesis techniques, which provide a difficult but promising
challenge for the materials synthesis laboratories.

\ack{Authors acknowledge financial support from the  project
PERMASOL (FFG project number: 848929).}

\end{document}